\documentclass[11pt]{article}
\pdfoutput=1 

\usepackage[no-natbib-sort]{jheppub} 
\usepackage[utf8]{inputenc}
\usepackage{amsfonts}
\usepackage{mathrsfs}
\usepackage{physics}
\usepackage{footnote}
\usepackage{scrextend}
\usepackage{leftidx}
\usepackage{soul}
\usepackage{makecell}
\usepackage{tabularx}
\usepackage{cprotect}

\setcitestyle{authoryear}
\renewcommand{\cite}[1]{\citep{#1}}

\usepackage[dvipsnames]{xcolor}
\definecolor{red}{rgb}{0.9, 0,0}
\definecolor{cerulean}{rgb}{0., 0.42,0.9}
\definecolor{prettygreen}{rgb}{0., 0.55,0.3}

\usepackage{epsfig}
\usepackage{graphicx}  
\usepackage{url}
\usepackage{color}
\hypersetup{
     colorlinks   = true,
     citecolor    = red,
	 linkcolor=blue
}
\usepackage{float}
\usepackage{placeins}

\def\beq{\begin{eqnarray}}
\def\eeq{\end{eqnarray}}
\def\bea{\begin{eqnarray}}
\def\eea{\end{eqnarray}}

\definecolor{red}{rgb}{0.9, 0,0}
\definecolor{cerulean}{rgb}{0., 0.5,0.8}

\graphicspath{{./figures/}}

\title{Snowmass2021 Theory Frontier: Theory Meets the Lab}
\author[a]{Rouven Essig}
\affiliation[a]{C.N. Yang Institute for Theoretical Physics, Stony Brook University, NY 11794, USA}

\author[b,c,d]{, Yonatan Kahn}

\affiliation[b]{Department  of  Physics,  University  of  Illinois  at  Urbana-Champaign,  Urbana,  IL  61801,  USA} 
\affiliation[c]{Illinois  Center  for  Advanced  Studies  of  the  Universe, University  of  Illinois  at  Urbana-Champaign,  Urbana,  IL  61801,  USA}
\affiliation[d]{Superconducting  Quantum  Materials  and  Systems  Center  (SQMS), Fermi  National  Accelerator  Laboratory,  Batavia,  IL  60510,  USA}

\author[e,f]{, Simon Knapen}
\affiliation[e]{Theoretical Physics Group, Lawrence Berkeley National Laboratory, Berkeley, CA 94720, USA}
\affiliation[f]{Berkeley Center for Theoretical Physics, Department of Physics, University of California, Berkeley, CA 94720, USA}

\author[g]{, Andreas Ringwald}
\affiliation[g]{Deutsches Elektronen-Synchrotron DESY, Notkestr. 85, 22607 Hamburg, Germany}
\author[i]{, and Natalia Toro}
\affiliation[i]{SLAC National Accelerator Laboratory, 2575 Sand Hill Road, Menlo Park, CA 94025, USA}

\date{March 2022}
\abstract{We review how theorists have been instrumental in launching and developing new experiments in the last decade, and comment on the challenges and opportunities for this program to continue to thrive going forward. This whitepaper is a solicited contribution to the Snowmass2021 Theory Frontier.}

\begin{document}

\maketitle
\beforetochook

\section*{Executive summary}
Theory and experiment have always advanced together, yet in the modern era of particle physics each has become a distinct specialty. It is notable, then, that the last decade has seen a surge of interest in small experiments where theorists have often played an integral part.
This pattern is especially evident in the three branches of the Dark Matter New Initiatives (DMNI) program and in the experiments searching for long-lived particles using LHC detectors far from the interaction points. This whitepaper examines the history and status of these fast-developing experimental programs, focusing on the time period since the previous Snowmass process, to address three questions: 
\begin{enumerate}
    \item[\emph{(i)}] Why has such great theoretical interest been focused on these new directions? 
    \item[\emph{(ii)}] How has close theoretical engagement catalyzed the launch of new experiments?
    \item[\emph{(iii)}] What lessons do the initiatives of the past decade hold for the future?
\end{enumerate}
Our discussion can be summarized as follows:

\emph{(i)}
The roots of theoretical interest in new experiments are closely tied to the impressive constraints on TeV-scale physics obtained by both the LHC and dark matter (DM) detectors. These results do not change the theoretical importance of exploring the TeV-scale and understanding the electroweak hierarchy, but they have placed pressure on the widespread assumption that the DM and  hierarchy (or strong CP) puzzles would be solved together.   Taken separately, each puzzle admits a slightly broader space of theoretical solutions, but with a \emph{vastly} broader range of observable signals. Many of these were ripe for exploration with small experiments.  Renewed appreciation of DM unrelated to TeV-scale puzzles --- but based on taking DM, in its own right, as strong and direct evidence for physics beyond the Standard Model --- called for exploring the much broader space of viable and detectable DM models.  Meanwhile, the ubiquity of unstable but long-lived particles in TeV-scale theories, untethered from DM, also prompted new kinds of experiments at the LHC beyond the standard collider detectors ATLAS and CMS.  \textbf{In short, new theoretical vistas have come into focus, spurring the identification of observables that are both theoretically compelling and different enough from established models that they call for new experimental directions.} 

\emph{(ii)}
\textbf{Theorists have played key roles in developing new experiments, from motivation through conception, design, and execution.} 
Theorists have identified new theoretical windows of interest, then synthesized the implications of existing data to these theories and the parameter-space milestones that are both compelling and plausibly accessible.  Today's theorists have acquired the necessary experimental literacy to find new conceptual pathways to discovery, leveraging not only traditional particle detectors but also innovations from fields as disparate as atomic physics, condensed matter physics, and quantum sensing. And theorists' calculations have often delivered a first preview of backgrounds and analysis challenges that new styles of experiments will face, informing the optimization and design of new experiments and the subsequent analysis of their data. These roles of theorists are particularly vital in experimental fields that are still emerging, where experts in instrumentation are still students of the new scientific areas their experiments can explore. 


We explore the various facets of theorists' roles in developing new experimental directions by focusing on three research areas where theorists have made transformative progress over the last decade: accelerator searches for DM, with both low-energy and high-energy beams; low-threshold direct detection of DM, and wave-like DM detection. For the first two areas, theorists played a pivotal role in creating new experimental fields, while in the third area, theorists have been essential to both motivating and enabling searches for axion DM outside the classical window. So far these efforts have been supported by a combination of DOE, NSF, national laboratories, private foundations, and foreign funding agencies, with the DOE Office of High Energy Physics's Dark Matter New Initiatives (DMNI) program playing a key role.


\emph{(iii)}
Looking further ahead, it is clear that the experiments we focus on here are only the first step towards a broader and more diversified experimental landscape in particle physics. 
Two main challenges remain:
\begin{itemize}
    \item There is a great need for \textbf{more for cross-talk between theory and instrumentation experts, as well as between theorists across different fields of physics} (for example, particle physics, condensed matter, and AMO), who are often outside each other's existing professional networks. Supporting efforts to forge these connections is therefore essential, as it directly benefits the search for dark matter and new particles.
    \item A \textbf{stable, continuing source of funding} for small, innovative experiments  and associated R\&D is needed to enable the realization of these ideas; the next generation of particle physics experiments may have its seeds in small-scale projects waiting for funding today.  
\end{itemize} 
Continuing the current diversification of our experimental program is critical for the future of particle physics, as it is less clear than ever where the next sign of new physics will raise its head. Without a strong theory community, this effort would not only lack the ability to judiciously set priorities, but would likely also miss out on new, potentially ground-breaking experimental concepts.

 \clearpage
\beforetochook\hrule
\tableofcontents
\afterTocSpace
\hrule
\afterTocRuleSpace
\pagestyle{myplain}

\section{Introduction}
\label{sec:Introduction}
Progress in fundamental physics has always been fueled by a close interplay between theory and experiment \cite{pais1986inward}. 
Theory advances are essential to interpret the observations made in our laboratories. The discovery of the Standard Model (SM), the crowning achievement of particle physics in the twentieth century, hinged on this interplay between discovering new phenomena and unraveling their underlying principles. 
Theory is also crucial to identify inconsistencies between experimental results and to propose their resolution, both of which further new experimental directions. The most spectacular example of this in recent history was the 2012 discovery of the Higgs boson \cite{ATLAS:2012yve,CMS:2012qbp}, whose existence was predicted back in 1964 \cite{PhysRevLett.13.321,HIGGS1964132}. The construction of the Large Hadron Collider (LHC) and the design of its detectors were heavily influenced by the predicted properties of the Higgs, and its discovery was a true triumph for both experimental and theoretical particle physics. 

In addition to the LHC, twenty-first century physics also hosts a vibrant program to detect dark matter (DM). 
Modern, large-scale DM experiments trace their origins in part to theory work that pointed out the utility of nuclear recoils \cite{PhysRevD.31.3059,Drukier:1986tm} and high-quality resonant cavities \cite{Sikivie:1983ip}.
These investigations have become even more important as the gravitational evidence for DM has accumulated inexorably, from the rotation-curve measurements \cite{1970ApJ...159..379R} which were already available at the dawn of the direct-detection era to the recent precision measurements of the Planck satellite \cite{Aghanim:2018eyx}, which can determine the cosmic DM density to sub-percent-level accuracy. Indeed, \textbf{it may be persuasively argued that DM is the strongest experimental evidence for the incompleteness of the Standard Model.}

The importance of identifying the particle nature of DM has motivated an ever-growing spectrum of approaches and detector technologies to search for DM candidates with a very wide range of masses and properties. 
The emphasis on broadening the search for DM, to maximize our chances of discovery, reflects a significant shift in perspective within the theory community.  
Outstanding puzzles such as the hierarchy problem and the strong CP problem continue to drive both model-building and experiments at the energy and precision frontiers. At the same time, impressive constraints from the LHC and direct detection experiments have driven theorists to critically re-examine theory priors that tied DM to these puzzles' resolutions.  Greater appreciation of the broad viable mass range for thermally-produced DM, the non-uniqueness of post-inflationary axion production, and the ubiquity of light bosons in string compactifications has brought new motivations, possibilities, and discovery opportunities for DM into focus. In parallel, the realization that many theories of beyond-the-Standard Model (BSM) physics contain metastable particles with production, decay, or interaction properties beyond the reach of standard multi-purpose collider detectors has inspired new concepts for auxilliary LHC detectors.

As a result of this evolution, a significant fraction of the theory community is now deeply invested in identifying new experimental probes for BSM physics, especially in the search for signals to which existing experiments have little or no sensitivity. This \textbf{``theory to the lab''} pipeline has proven to be a rather \textbf{multi-disciplinary} endeavor, as the latest innovations in atomic physics, condensed matter physics, quantum sensing, nuclear physics, astrophysics, and cosmology are actively sought out and studied with an eye toward possible particle physics applications. We have also seen a much increased \textbf{experimental literacy} amongst theorists, as many have acquired a detailed understanding of existing and upcoming detector technologies, resulting in creative alternative uses for detectors originally designed with a different purpose in mind (for example, the observation that silicon photodetectors can sense DM-electron scattering). The role of theorists in experimental collaborations has also changed, with members of the theory community running the gamut between generating enthusiasm for an initial idea to serving as co-PIs of a dedicated experimental effort. These new paths for theorists have opened up opportunities for collaboration with experimentalists (from all fields of physics, not just particle physics!) at all stages of experimental development, from conception to execution. For all these reasons, theorists have been crucial in launching several qualitatively new experimental directions over the past decade.


The remainder of this white paper is devoted to telling the stories of the development of several types of experiments representative of progress over the past decade in three directions: accelerator-based experiments (Sec.~\ref{sec:Accelerator}), particle-like DM detection (Sec.~\ref{sec:ParticleDM}), and wave-like DM detection (Sec.~\ref{sec:AxionDM}). These three probes were identified by a recent DOE Basic Research Needs (BRN) study on Dark Matter New Initiatives (DMNI) \cite{BRN} as Priority Research Directions for the field over the next decade. Our goal here is to highlight and emphasize the role of theorists in conceiving, developing, and executing these experiments, and to show how these new trails blazed can guide the way forward for the next generation of theorists to continue this important effort (Sec.~\ref{sec:RoadAhead}). Here an important caveat is in order: we make no effort to give an exhaustive survey of all new experiments, and the choices we have made reflect the authors' personal familiarity with the experiments in question. We feel that such a structure will allow the small details that illustrate the collaborative, personal, and often stochastic nature of scientific research to shine forth.

\section{New directions in the theory of dark matter}
\label{sec:theory}

In this section, we briefly review some of the main features of the DM-inspired models outlined in the introduction, with the intent of establishing some terminology and providing some minimal context and references. We will postpone discussion of TeV-scale models, including those related to the hierarchy problem, to Sec.~\ref{sec:LHC}, where they are most applicable. 
We note that a systematic exploration of the parameter space of DM-inspired models began in earnest during the last Snowmass process \cite{Essig:2013lka}, including its lead-up~\cite{Proceedings:2012ulb}, and the subsequent Dark Sectors \cite{Alexander:2016aln}, Cosmic Visions: New Ideas in Dark Matter \cite{Battaglieri:2017aum}, Dark Matter New Initiatives \cite{BRN}, and Feebly Interacting Particles (FIPs) \cite{Agrawal:2021dbo} workshops; in this whitepaper, we wish to emphasize the enormous progress in new experiments probing this parameter space in just the past decade.

Pragmatic ``bottom-up'' theory priors aim to cover the widest possible range of SM extensions that \emph{i)} are theoretically consistent, \emph{ii)} do not conflict with existing measurements, and \emph{iii)} could conceivably be discovered in the next decade with existing or upcoming technology. The combination of these three criteria is surprisingly restrictive, and provides a useful blueprint for progress  that is complementary to the traditional, more top-down line of thinking. 
An important benchmark in the space of DM-inspired models is the concept of a \textbf{dark sector}. It is most powerful for low-mass ($\lesssim 1$ GeV) extensions of the SM: \emph{If} a new sector has a typical mass scale of a few tens of GeV or lower, it \emph{must} be weakly coupled to the SM \cite{Essig:2013lka}, hence the name dark sector. In particular, this framework excludes particles that are multiplets under the SM gauge groups\footnote{``True millicharged'' states with hypercharge $\ll1$ are phenomenologically allowed, though they obstruct grand unification.  These have similar phenomenology to the ``effectively'' millicharged particles from a dark sector, discussed further below.}, such as most supersymmetric partners or canonical WIMP DM. 

An obvious interpretation of some of the particles in the dark sector is that they could comprise DM candidates; the Lee-Weinberg bound \cite{Lee:1977ua} on thermal freeze-out of light particles through weak interations then implies that sub-GeV DM must interact with a new light mediator \cite{Boehm:2003hm,Boehm:2003ha}, which would also be part of the dark sector. If that mediator were a massive gauge field, it could obtain its mass through spontaneous gauge symmetry breaking by a new scalar field, yet another possible component of the dark sector.
One can systematically study interactions between the dark sector and the SM by classifying such interactions according to the Lorentz representation of the new particle(s). In particular, the simplest allowed renormalizable couplings between the SM and a light dark sector are mixing of the photon, SM Higgs boson, or neutrinos with a new neutral particle of the same spin. Each of these interactions, known as \textbf{portals}, motivates a distinct and predictive pattern of interactions between the dark particle and ordinary matter. These form a basic framework for searching for low-mass new physics, including the DM, the mediator itself, and other accompanying light states. 

One of the earliest portals considered is the so-called \textbf{dark photon} \cite{Holdom:1985ag} (a new vector mixing with the SM photon), which is now a very common benchmark for experimental searches, a useful ingredient in many UV-complete models, and a viable mediator through which a DM candidate could couple to the SM. Simple models of DM coupled through a dark photon span a much wider range of masses and couplings than the traditional WIMP \cite{Pospelov:2007mp}. In the limit where the dark photon mass is very small, this model moreover reproduces the phenomenology of a \textbf{millicharged particle}, without spoiling grand unification. Variations where one gauges an anomaly-free subgroup of the SM flavor group, e.g.~$L_\mu-L_\tau$ are also popular, in particular in the context of the muon $g-2$ measurements \cite{PhysRevLett.126.141801}. One may also extend the SM with a single, real scalar field that can mix with the SM Higgs. Such a \textbf{dark Higgs} can exist on its own or be a mediator to a more complicated dark sector. It can be produced in, for example, exotic Higgs or exotic $B$ decays \cite{Willey:1982ti,Chivukula:1988lo,Grinstein:1988yu,Curtin:2013fra}. \textbf{Axion-like particles (ALPs)} are pseudo-scalars that interact with the SM matter and/or gauge fields, with possible interactions restricted to shift-symmetric operators due to the pseudo-Goldstone nature of the axion. Originally proposed to solve the strong CP problem \cite{Peccei:1977hh,Peccei:1977ur}, ALPs are also ubiquitous low-energy remnants of string theory and leave observable signatures in cosmological, astrophysical, direct detection, and collider probes \cite{Irastorza:2018dyq}. We can also extend the SM with a (set of) new fermions, which mix with the Standard Model neutrinos. These \textbf{heavy neutral leptons} are motivated in mechanisms to generate the neutrino masses \cite{Mohapatra:1986bd} and/or the baryon asymmetry in the Universe \cite{Asaka:2005an,Asaka:2005pn}.

In the dark sector framework, DM may interact with the SM by coupling directly to one of the portal particles. Its interactions may also be classified by its Lorentz representation: in the case of a gauge coupling, DM may be a scalar, pseudoscalar, Majorana fermion, or Dirac fermion. For thermal DM, the latter possibility is very tightly constrained by the absence of late-time annihilation as seen in CMB data \cite{Slatyer:2015jla}. Alternatively, the DM itself may be one of the portal particles, such as a dark photon or an axion. In this case, the correct DM relic density may be obtained through non-thermal processes such as the misalignment mechanism \cite{Preskill:1982cy,Abbott:1982af,Dine:1982ah} or gravitational production during inflation \cite{Graham:2015rva}. Obtaining the correct relic abundance while evading astrophysical and cosmological constraints generically requires such DM candidates to be \textbf{ultralight}, with mass much less than 1 eV, in which case they have macroscopic phase space occupancy today. In this sense, such DM acts more like a coherent field than individual particles, leading to qualitatively different detection strategies (often relying on precision metrology) compared to traditional DM searches. 

Taking the mile-high view of the DM landscape, the union of dark-sector DM with ultralight DM candidates is bounded from above by the $\simeq 1 \ {\rm GeV}$ limits of traditional WIMP DM experiments, and from below by small-scale structure constraints at $\simeq 10^{-22} \ {\rm eV}$ \cite{Ferreira:2020fam}. This yields a \textbf{DM parameter space covering at least 31 orders of magnitude}, where models exist at nearly every possible mass that feature a consistent thermal history and no conflicts with any terrestrial or astrophysical observations.\footnote{While we do not focus on them in this review, there is also renewed interest in heavy DM with purely gravitational couplings \cite{Kolb:1998ki,Blanco:2021yiy}, involving close collaboration between theorists and experimentalists \cite{Carney:2019pza,Carney:2020xol}. Including such DM candidates widens the possible mass range of DM to more than 50 orders of magnitude.} Clearly, there is no one-size-fits-all solution for the entire space of DM models, necessitating the development of many targeted experiments to cover all possibilities. A natural dividing line is the eV scale, above which DM is sparse in phase space and may be considered \textbf{particle-like}, and below which DM has macroscopic phase space occupancy and is \textbf{wave-like}.

Finally, non-minimal dark sectors with rich internal structure are often referred to as \textbf{hidden valleys}. These may include confining dynamics within the dark sector, leading to distinctive collider phenomenology \cite{Strassler:2006im}.  While they are valued as purely as bottom-up extensions of the SM, they also naturally occur in the models of asymmetric DM \cite{Kaplan:2009ag} and neutral naturalness \cite{Craig:2014aea,Craig:2015pha,Curtin:2015fna}. We will return to the latter in detail in Sec.~\ref{sec:LHC}. We also note that many of the dark sector components we have identified are ubiquitous low-energy remnants of the string theory landscape \cite{Halverson:2018vbo}, providing an important phenomenological link between bottom-up and top-down reasoning.

\section{Accelerator-based experiments}
\label{sec:Accelerator}

The last decade has seen a new appreciation of accelerator-based experiments as powerful and direct windows into new physics.  This realization required a broadening of theoretical focus --- expanding from TeV-scale physics with ${\cal O}(1)$ couplings to ordinary matter, such as WIMPs and weak-scale supersymmetry, to encompass the dark sectors described in Sec.~\ref{sec:theory} above. Exploring dark sectors also called for new approaches to experimental searches, many of which were first proposed by theorists. In particular, low-energy \textbf{fixed-target experiments} offered the high luminosities needed to probe very weak couplings and called for compact forward detectors that could be built at low cost. Many were so-called ``parasitic'' experiments, highly leveraging existing capabilities and equipment, and in some cases running in the background without dedicated beam-time.  Theorists were, in many cases, instrumental to the development and optimization of these new experiments and worked closely with experimentalists to realize them. 

Though the ideas of dark sectors, compact forward-facing experiments, and the parasitic approach were initially pulling attention away from the energy frontier, they have now turned full-circle and informed new ways of exploiting LHC collisions. Theoretical ideas --- including the realizations that low-$p_T$ LHC collisions could copiously produce low-mass dark sector particles and that heavier long-lived particles are ubiquitous in TeV-scale physics models --- were once again key to unlocking these opportunities. And theorists again played leading or key roles in developing new experiments that exploited these insights. 



\subsection{Low energy accelerators}
Fixed-target experiments and low-energy colliders have a rich history in the early development of the Standard Model (for example, the $b$ quark was discovered through production of the $\Upsilon$ resonance in a 1977 fixed target experiment at Fermilab \cite{Herb:1977ek}, while the $\tau$ lepton and $c$ quark were discovered in few-GeV colliders), but since the 1990s the main focus of particle physics searches was on colliders energetic enough to explore the electroweak scale and beyond, such as LEP, the Tevatron, and the LHC.  Modern facilities and experiments in the few-GeV range have mainly been developed either for \emph{indirect} probes of high-energy physics (e.g. flavor physics, M\"oller scattering); for nuclear and hadronic physics; or for production of neutrino beams.  

A resurgence of interest in low-energy experiments as particle-discovery machines emerged from the rise of interest in exploring the possibility of dark sectors. Much of this interest was initially spurred by anomalies potentially related to DM (e.g. \cite{Boehm:2003ha,pospelov:2008jd,Arkani-Hamed:2008hhe}), but evolved into the aim of exploring the dark-photon portal, and dark sectors more generally, as broadly as possible. The first step in this program was identifying new applications of existing data. These included recasting decades-old beam-dump searches for visible axions (e.g. \cite{Bjorken:2009mm, Batell:2014mga}) and the LSND neutrino oscillation experiment (\cite{Batell:2009di,deNiverville:2011it}) as constraints on dark photon and light DM production, and proposing new analyses of B-factory and meson-decay data (see e.g. \cite{Borodatchenkova:2005ct,Reece:2009un, Batell:2009jf, Essig:2009nc,Essig:2010gu, Essig:2013vha}). 

These studies made clear that very interesting parameter space was accessible to intensity-frontier experiments in principle, but beyond the reach of present data.  The next logical step, therefore, was identifying small-scale experiments that could improve on the sensitivity of their predecessors.

An early example was \cite{Bjorken:2009mm}, a paper by four theorists that assessed existing constraints on the minimal dark photon model and went on to propose several new experimental concepts to extend this sensitivity. These concepts were all fixed-target experiments using an electron beam, but with different detector configurations tailored to the event rate and lifetime expected in different parts of the dark photon parameter space. Two of these concepts have since been developed into successful experiments at Jefferson Lab (JLab)'s CEBAF 6 GeV electron accelerator. \textbf{APEX} \cite{Essig:2010xa, APEX:2011dww} configured the high-resolution spectrometers in JLab's Hall A as a pair spectrometer operating at high rate to search for dark photons at couplings just below those accessible to colliding-beam experiments.  This ``rapid response'' experiment (the first run occurred and was published within 20 months of the proposal) was led by a team of one experimentalist and three theorists.
This work also inspired a similar experiment at the Mainz Microtron \cite{A1:2011yso}. 
Around the same time, the same theory collaborators teamed with SLAC and JLab experimental teams to develop a compact, forward silicon-based spectrometer concept optimized to look for the displaced decays of dark photons with even lower couplings, which became the \textbf{Heavy Photon Search (HPS)} experiment \cite{Battaglieri:2014hga,HPS:2018xkw}.  

Theorists' core involvement in these experiments was important on multiple levels. First, they offered the conceptual seed for these experiments' designs, including early glimpses of the backgrounds and design trade-offs that they would face. Second, they had the necessary familiarity with the theoretical and experimental landscape, including past results from experiments in several disparate research areas. This allowed them to anticipate the sensitivities of re-analyses that had not yet been done, but could be completed well before the new proposals would come online. These experiments' development benefited greatly from encounters that brought the theory teams and interested experimentalists together, such as the SLAC 2009 dark forces workshop and visits of several of the authors of \cite{Bjorken:2009mm} to JLab.  

A second thrust in low-energy searches for dark sectors focused on the production of DM.  These experiments take two main strategies.  The first, pioneered in \cite{deNiverville:2011it}, was production of DM in collisions of a very intense beam against a thick target or beam dump, followed by its detection in downstream detectors. After applying this logic to LSND, the authors recognized that it could be powerfully exploited at MiniBooNE as well. These theorists and their collaborators were instrumental in the proposal for a dedicated MiniBooNE run with the beam configured to \emph{miss} the neutrino-production target \cite{MiniBooNE:2017nqe, MiniBooNEDM:2018cxm}, which minimized neutrino backgrounds in a DM search, as well as in the development of the \textbf{Coherent CAPTAIN-Mills} DM search program \cite{Aguilar-Arevalo:2021sbh}.  The broad idea that accelerator-based neutrino experiments can also search for dark-matter production has now been widely studied and applied, as surveyed in \cite{Buonocore:2019esg}.  Theorists were also involved in developing the related electron beam-dump concept \cite{Izaguirre:2013uxa}, realized through the \textbf{BDX} proposal at JLab \cite{BDX:2014pkr}, which involves a compact, dedicated detector that operates parasitically down-stream of a high-current electron beam dump used for other experiments (in contrast with the proton-beam concepts, which re-purpose a large neutrino detector but benefit from dedicated beam time).

From the beginning, dark-matter search experiments have found motivation in the fact that simple mechanisms for thermal freeze-out of DM tie the (unknown) portal coupling to the DM annihilation cross-section in the early Universe (inferred from the DM density).  However, the problem has multiple unknowns --- besides the portal coupling relevant for both cosmology and direct production, the cross-section depends on the DM and mediator masses and the dark-sector gauge coupling.  In \cite{Izaguirre:2015yja} it was appreciated that a fairly conservative \emph{lower bound} on production rate compatible with this simple freeze-out mechanism can be derived by assuming a dark-sector gauge coupling near its perturbativity limit and a small hierarchy between dark sector masses.  This logic motivated a ``thermal DM target'' as a natural goalpost, just a few orders of magnitude beyond currently proposed experiments' capabilities. 
This motivated new conceptual strategies: the electron-beam missing-energy strategy  \cite{Andreas:2013lya} exemplified by \textbf{NA64} leveraged the unusual dark-sector production kinematics highlighted in \cite{Bjorken:2009mm} to identify DM \emph{production events} from the substantial energy carried away by invisible DM particles, without detecting their interactions at all.  The missing-momentum approach \cite{Izaguirre:2014bca} extended this idea by adding a tracking detector, strongly motivated by the goal of rejecting neutrino backgrounds to the level required to reach the thermal DM milestones noted above. Theorists involved in developing the missing-momentum approach were also instrumental in building the \textbf{LDMX} collaboration \cite{LDMX:2018cma}, which has designed a high-rate-capable missing momentum detector and is funded through the planning stage by the DOE DMNI program. A similar experiment using a muon beam, \textbf{NA64-$\boldsymbol{\mu}$}, which would be able to test models of freeze-out such as $L_\mu - L_\tau$ which do not couple to first-generation SM particles at tree-level, is expected to take first data shortly \cite{Sieber:2021fue}. 

\subsection{LHC based detectors\label{sec:LHC}}
Particles with macroscopic lifetimes have always been a fertile ground for expected and unexpected discoveries, such as the muon, kaons, and $B$-mesons, just to name a few. As the energy of the collider and the masses of the particles under study got higher and higher, so did the naive estimates of the widths of these particles. The Higgs, top quark, $W$ and $Z$ all followed this trend by decaying promptly, as was anticipated. 

Most leading paradigms for BSM physics, however, predict one or more new long-lived particles (LLP's). One of the oldest and most famous examples may be the disappearing track signature in anomaly mediated supersymmetry \cite{Giudice:1998xp,Randall:1998uk,Feng:1999fu}, though macroscopic displacements are also expected in split supersymmetry \cite{Arkani-Hamed:2004ymt}, superWIMPs \cite{Feng:2003xh}, models with light gravitinos \cite{CABIBBO1981155} such as gauge mediation \cite{Dimopoulos:1996vz,Dine:1994vc} and models involving R-parity violation \cite{Barry:2013nva}. Of course, the LLP itself cannot be the DM if its decay can be observed in a collider experiment, but DM can be among the decay products in some of the above examples. This scenario is realized explicitly in freeze-in models, where the macroscopic decay lengths are required to set the correct DM relic density \cite{Hall:2009bx}. If the mechanism responsible for the neutrino masses takes place collider accessible energy scales, it may moreover produce heavy neutral leptons at colliders, which also decay with a macroscopic displacement \cite{Gorbunov:2007ak}. What all these models have in common is that the macroscopic lifetimes are the result of either a separation of scales or an approximate symmetry that is broken by a small parameter. Conversely, models with these features \emph{generically} predict LLPs.

The latter point especially was well appreciated amongst the part of the theory community working on ``dark sector'' models, which often follow a somewhat more bottom-up philosophy, as described in Sec.~\ref{sec:theory}.
While these ideas were appreciated at the time, the main focus of the theory and experimental communities during the start-up phase of the LHC was still on finding the Higgs as well as traditional solutions to the hierarchy problem, predominantly supersymmetry. Given this timing, the proposal for the \textbf{SHiP} experiment at the CERN SPS  \cite{Bonivento:2013jag} was truly remarkable in both terms of its ambition and science targets: it was the first time that such a large scale experiment was envisioned specifically with the purpose of looking for exotic long-lived particles. Its primary motivation was to go after the extensions of the neutrino sector described above, but its scope was quickly broadened to include other hidden sector models \cite{Alekhin:2015byh}. SHiP quickly grew into a large experimental collaboration with a very vibrant R\&D effort \cite{SHiP:2015vad}.  From its inception, the SHiP proposal was a joint venture of theoretical and experimental physicists.

As part of the bottom-up philosophy, particles with very small electric charges are quite compelling because of their theoretical simplicity and very interesting and non-trivial phenomenology. (Such new particles are usually called ``millicharged particles'', regardless of their precise charge, as long as it is  $\ll1$.) In particular, for charges $\lesssim 0.1$, they are below ionization thresholds of the silicon trackers in ATLAS, CMS and LHCb, and remain invisible to these detectors. This inspired a group of theorists and experimentalists to propose the \textbf{Miliqan} experiment \cite{Haas:2014dda}. The key idea is that suitable set of large scintillator bars can detect charges much smaller than unity, while eliminating backgrounds by a combination of shielding, light yield measurements, and coincidence constraints between different parts of the detector. The required shielding can be achieved by placing the experiment in drainage gallery above the CMS experiment, which is the equivalent of 17 m of rock. Miliqan has already taken data with a demonstrator detector, and its main detector is fully funded, under construction, and on track to take data during run 3 of the LHC \cite{milliQan:2021lne}. Miliqan's sister experiment, called \textbf{SUBMET}, \cite{Kim:2021eix}, is being constructed at J-PARC.

The bottom-up line of thinking was also instrumental in amplifying and understanding the potential of \textbf{LHCb} as an experiment ideally suited for dark sector searches. While LHCb had been performing a number of searches for light, beyond-the-standard model particles \cite{LHCb:2014jgs,LHCb:2015nkv}, it was the joint work between the MIT theory and LHCb groups which brought LHCb's online capabilities to the attention of the broader theory community \cite{Ilten:2015hya,Ilten:2016tkc}. This kick-started a broad and fruitful collaboration between BSM theorists and LHCb experimentalists, which explored the complementarity between LHCb, ATLAS and CMS in the search for dark sector particles \cite{Borsato:2021aum}.

The most forward-looking effort to discover light new particles at the LHC is probably the \textbf{FASER} experiment \cite{FASER:2018bac}, which sits 480 m meters downstream from the ATLAS interaction point. The key ideas behind FASER are the huge flux of particles that are produced at high rapidity and the large amount of rock that is shielding the detector from direct backgrounds from the interaction point. FASER is unique among the examples in this subsection, in the sense that it was first proposed by a group of \emph{theorists only} \cite{Feng:2017uoz}.  Of course the design has evolved since the original theory paper, but the core idea was viable and experimentally realistic from the very beginning. This was possible due to the high level of experimental sophistication among this group of theorists, something which is increasingly common in today's theory community and a major contributor to progress in this field. FASER is now an established, vibrant, and funded experimental collaboration and the installation of its first detectors was completed in March 2021. First physics results are expected in the coming years. In addition to searches for dark sector particles, the FASER setup can also be used to probe TeV-scale neutrinos \cite{FASER:2019dxq}. 

So far we have explored the consequences of the mostly bottom-up method of extending the Standard Model\footnote{See \cite{Beacham:2019nyx} and \cite{Agrawal:2021dbo} for detailed reviews on the experimental implications of this line of thinking.} for LHCb as well as some of the new experiments such as SHiP, Miliqan, and FASER. During the early phase of the LHC these developments could not be taken for granted however, and most of the theory community was still focused on probing traditional, top-down naturalness paradigms, especially supersymmetry. This changed radically around 2015, when it became clear that the stop squark was not going to be found in the first 20~$\text{fb}^{-1}$ of data from ATLAS and CMS. The ``Neutral Naturalness'' paradigm took root around this time, as it could explain the stability of the weak scale without colored matter at TeV energies. This class of models was based on the old Twin Higgs \cite{Chacko:2005pe} and Folded Supersymmetry models \cite{Burdman:2006tz}, but only in 2015 it was realized that LLP's are \emph{generic} predictions in this set of models \cite{Craig:2015pha,Curtin:2015fna}. A vibrant community of theorists and experimentalists working on LLPs was formed around this time, cumulating in an extensive review of the LLP landscape \cite{Alimena:2019zri}.

In particular, in April of 2016 a theory workshop for this purpose was assembled at the University of Maryland, which also included a handful of experimentalists with expertise on searches for LLPs. The discussions focused heavily on identifying better ways to detect the LLP's with very long lifetimes that were predicted by Neutral Naturalness models. As the main figure of merit is the size of the fiducial detector volume, initial (rather  ambitious) ideas included a fleet of balloon- or drone-borne detectors above the Meyrin site at CERN, or instrumenting parts of Lac Léman. The eventual \textbf{MATHUSLA} proposal \cite{Chou:2016lxi} was published several months later by a collaboration of theorists and experimentalists that had all attended the workshop. In this sense, a theory workshop had directly let to the birth of a new experimental collaboration because the organizers had invited a handful of adventurous experimentalists. 

The MATHUSLA proposal \cite{MATHUSLA:2018bqv} envisions a large detector to be installed on the surface above the CMS interaction point, to catch LLP's decaying in flight. The rock between the CMS interaction point and the surface detector provides the necessary shielding. For MATHUSLA to work, it must be very large (100m $\times$ 100m $\times$ 30m in the current design \cite{MATHUSLA:2020uve}) and an affordable yet reliable detector technology is needed. Originally Resistive Plate Chamber panels were envisioned, but currently the collaboration is aiming for cheaper plastic scintillator technology. The current priorities of the collaboration are finalizing the technical design report for the full experiment and the construction of a prototype detector.

The idea for the \textbf{CODEX-b} experiment \cite{Gligorov:2017nwh} originated from discussion between theorists and LHCb experimentalists in the CERN cafeteria. It was inspired by the concept of MATHUSLA and the roughly $(10\,\text{m})^3$ space that would open up in LHCb cavern after their DAQ systems would be relocated to the surface. The detector intends to be less powerful than the nominal MATHUSLA design but comes at much lower cost.  The CODEX-b collaboration \cite{Aielli:2019ivi,Aielli:2022awh} has accreted a healthy number of experimentalists, who are working on the design of the experiment, detailed background simulations, and the commissioning of their $(2\,\text{m})^3$ demonstrator detector. The demonstrator is fully funded and the collaboration is aiming for its installation in 2022--2023.

We would be remiss to not briefly comment on a number of other, similar proposals and experimental efforts at the LHC. The \textbf{MOEDAL} experiment makes use of the LHCb interaction point and is optimized to look for magnetic monopoles \cite{MoEDAL:2014ttp}. The collaboration is planning an extension called MOEDAL-MAPP in a forward service tunnel, at a rapidity between that of CODEX-b and FASER \cite{Frank:2019pgk}. The \textbf{ANUBIS} proposal targets the same parameter space as CODEX-b and MATHUSLA and would be suspended in the access shaft above CMS \cite{Bauer:2019vqk}. The \textbf{FACET} proposal \cite{Cerci:2021nlb} resembles FASER, but would be installed around the LHC beam pipe, downstream from CMS, rather than in a service tunnel. Finally, the \textbf{AL3X} thought-experiment was extremely ambitious and envisioned repurposing the ALICE infrastructure as an LLP detector after the projected lifetime of the experiment \cite{Gligorov:2018vkc}. At this time, there is no obvious path forward for AL3X, though its background studies are valuable for CODEX-b, MATHUSLA, MOEDAL-MAPP, and ANUBIS. 

\section{Dark matter direct detection below the WIMP mass}
\label{sec:ParticleDM}
The broadening of theory priors beyond the mass ranges preferred by the traditional WIMP and axion DM candidates has spurred significant innovation in both theory and experiment. Similarly to the accelerator probes discussed in Sec.~\ref{sec:Accelerator}, direct-detection experiments play an essential and unique role in the search for dark-sector DM. For example, while electroweak WIMPS have only contact interactions with nuclei, the possibility of a light mediator in the dark sector opens channels for momentum-dependent DM-matter interactions, which are enhanced at low momentum transfer and thus easier to detect for extremely light DM candidates, down to the keV warm DM limit. This is a double-edged sword, however: 
for sub-GeV DM interacting with nuclei, very little of the DM's kinetic energy is transferred to the energy of the recoiling nucleus, making such signals impossible to detect with conventional nuclear recoil searches as the DM mass is lowered. Even if the event rate is large, if the detection thresholds are too high, the signal remains invisible. Furthermore, it is crucial to account for the properties of the mediator, especially if the mediator has long-range interactions (however weak they might be) with the target constituents. The detailed, microscopic properties of the target material must therefore also be considered for when calculating the scattering rates. As a consequence of these constraints, the search for alternative signals that exploit lower-energy collective modes or quasiparticles in solid-state detectors has spurred a fruitful collaboration between particle theorists and condensed matter theorists~\cite{Kahn:2021ttr}. 

On the experimental side, new low-threshold sensors are being deployed or developed specifically for the purpose of searching for sub-GeV DM candidates. Rapid progress is possible, even with very small exposures, because the DM number density is orders of magnitude higher than for WIMPs (scaling inversely with the DM mass), and existing cross section bounds are rather weak. In this section, we discuss in some detail two examples of ideas proposed and worked out by theorists that have now evolved into very promising experiments that are already taking data: SENSEI and LAMPOST. We further briefly mention some more recent ideas that are currently still in the R\&D stage, but which are expected to have preliminary data within the next 5 years.

\subsection{MeV--GeV DM-electron scattering}
\label{sec:DMscattering}
Around 2010, theorists realized that several inelastic processes can allow for the transfer of most of the sub-GeV DM's kinetic energy to a target detector material~\cite{Essig:2011nj}.  An important example of this is DM-electron scattering in atoms or semiconductors; since the recoiling electron can ionize other atoms (in, e.g., noble liquid targets) or create additional electron-hole pairs (in, e.g., a semiconductor), the resulting signal consists of one or a few electrons (or electron-hole pairs).  When this was first proposed, it was not clear whether such small signals could ever be detected and, even if they could, whether they could be distinguished from backgrounds.  However, the theorists urged and joined a few eager experimentalists to look at data taken by \textbf{XENON10}~\cite{Angle:2011th} and were able to set the first direct-detection constraints on sub-GeV DM~\cite{Essig:2012yx}, down to masses of a few MeV.  At the same time, since the XENON10 data suffered from large backgrounds, the theorists also engaged with experimentalists working on \textbf{DAMIC} and \textbf{SuperCDMS} (using silicon or germanium detectors) to determine whether their sensor technology could be pushed to detect one- or few-electron events, which would allow one to probe DM scattering as light as $\sim$500~keV (and DM absorption down to the silicon band gap of $\sim$1.2~eV).  

Motivated by the strong theoretical reasons to search for sub-GeV DM and by the new detection concepts, a few experimentalists designed and improved their sensors to probe lower thresholds.  The theorists joined a Fermilab LDRD with the goal of developing an improved \textbf{``Skipper''} readout system for Charge Coupled Devices (CCDs), building on the success of the DAMIC CCD program (the sensors are designed by the Lawrence Berkeley National Laboratory (LBNL) and fabricated at Teledyne/DALSA).  While the theorists worked on improving the theoretical calculations for DM-electron scattering and sharpening the theory motivation~\cite{Graham:2012su,Lee:2015qva,Essig:2015cda}, the experimentalists on the LDRD, with great ingenuity, turned it into a remarkable success that led in 2017 to the first silicon detector able to sense single electrons with sub-electron noise precision~\cite{Tiffenberg:2017aac}.  Shortly thereafter, in another experimental feat of ingenuity, the high-voltage charge amplification with transition edge sensor readout \textbf{(HVeV)} technology from SuperCDMS was demonstrated to have single-charge resolution~\cite{Romani:2017iwi}.  These breakthroughs meant that DM scattering could now be detected down to the MeV scale.  

A proposal to build the \textbf{SENSEI} experiment, consisting of $\mathcal{O}$(100) grams of silicon Skipper-CCDs, was funded, with theorists and experimentalists jointly leading the collaboration and working together.  SENSEI has to date published three sub-GeV DM constraints and currently sets the strongest constraints on DM-electron scattering through a light mediator~\cite{Crisler:2018gci,Abramoff:2019dfb,SENSEI:2020dpa}, approaching the benchmark parameter space for non-thermal production through the freeze-in mechanism~\cite{Hall:2009bx,Essig:2011nj,Chu:2011be,Dvorkin:2019zdi}. The HVeV technology led to similar constraints on sub-GeV DM~\cite{Agnese:2018col,Amaral:2020ryn}.  Constraints have also been set by the noble liquid detectors \textbf{XENON100}~\cite{Aprile:2016wwo,Essig:2017kqs}, \textbf{DarkSide-50} (on which theorists and experimentalists again joined forces~\cite{Agnes:2018oej}), \textbf{XENON1T}~\cite{Aprile:2019xxb,XENON:2021myl}, and \textbf{PandaX-II}~\cite{PandaX-II:2021nsg}, as well as the solid-state detectors \textbf{DAMIC at SNOLAB}~\cite{Aguilar-Arevalo:2019wdi} and \textbf{EDELWEISS}~\cite{Arnaud:2020svb}. 
A larger experiment, \textbf{DAMIC-M}, with the design goal of a 1~kg Skipper-CCD detector, is also funded~\cite{Castello-Mor:2020jhd}.  Finally, with the involvement of particle and condensed matter theorists, an ambitious 10~kg Skipper-CCD detector called \textbf{Oscura} has received R\&D funding from the DOE DMNI program~\cite{Aguilar-Arevalo:2022kqd}.  

At the same time, to mention only a very few examples, theorists have developed numerous other novel DM detection ideas that take advantage of these technological advances to probe DM absorption~\cite{An:2013yua}, DM-nucleus interactions~\cite{Ibe:2017yqa},  and the solar- and cosmic-ray-accelerated component of sub-GeV DM~\cite{An:2017ojc,Emken:2017hnp,Bringmann:2018cvk}. Other theoretical detection concepts that are being implemented experimentally include DM scattering in scintillating crystals ~\cite{Derenzo:2016fse} and aromatic organic compounds \cite{Blanco:2019lrf}, the latter of which was able to set limits competitive with the initial surface runs of SENSEI despite being dark rate-limited. Theorists are also heavily involved in determining novel backgrounds for low-threshold experiments~\cite{Du:2020ldo,Berghaus:2021wrp}. 

\subsection{eV-scale DM absorption}
\label{sec:eVDM}
A key consequence of considering DM with masses below the MeV scale is that in some models it can be detected through \emph{absorption}, rather than scattering. For this process to be allowed, the DM must be unstable, however for sub-MeV DM its lifetime can well exceed the age of the universe. Absorption processes have already been considered in the detectors discussed in Sec.~\ref{sec:DMscattering}, see e.g.~\cite{XENON:2020rca,SuperCDMS:2020ymb,SENSEI:2020dpa,EDELWEISS:2020fxc,CDEX:2019isc,DAMIC:2019dcn}.  This has led theorists to devise new detection strategies that are optimized for absorption processes. 

The experimentally most advanced idea along these lines is a high frequency version of the dielectric haloscope concept (see Sec.~\ref{sec:antennas}) through the usage of photonic materials. Dielectric haloscopes modify the photon propagation in the target through a periodic structure of alternating layers of dielectric material, which allows for resonant axion-photon or dark photon-photon conversion. A team of  theorists realized that this idea could be extended to the 0.1 eV to 10 eV DM mass range by making use of photonic materials consisting of layers of ultra-thin films \cite{Baryakhtar:2018doz}. The outgoing photon can furthermore be focused with a mirror onto a low threshold quantum sensor, such as a transition edge sensor (TES), microwave kinetic inductance detector (MKID), superconducting nanowire single-photon detector (SNSPD) or single photon avalanche diode (SPAD). As the experiment is small, inexpensive, and relies on existing technology, two experimental collaborations were rapidly established. The \textbf{LAMPOST} collaboration \cite{Chiles:2021gxk} uses the SNSPD technology, while the \textbf{MuDHI} collaboration \cite{Manenti:2021whp} has opted for a SPAD sensor. Both experiments are already operating and recently announced their first physics results, with competitive limits in the dark photon parameter space.

A possible alternative strategy is to resonantly excite the rotation and vibration modes in gaseous, polyatomic molecules \cite{Arvanitaki:2017nhi} (see also \cite{Essig:2019kfe}). This detector concept can be read out in a similar fashion to the LAMPOST/MuDHI design.





\subsection{Towards sub-eV thresholds}

The desire to cover \emph{all} parameter space available for DM to scatter off a target has motivated theorists to consider DM down to the keV scale. This mass has a natural interpretation as a parameter space boundary, for at least two reasons: thermally-produced DM lighter than this would be ``warm'' and therefore unable to form small-scale structures that are observed in the matter power spectrum, and fermionic DM lighter than this would have a Fermi velocity exceeding the escape velocities of dwarf galaxies. The typical kinetic energy of such DM is on the meV scale, and thus our expanded theory priors provide strong motivation to push possible detection thresholds down to this scale. Moreover, the typical momentum of DM in this mass range is below the inverse interatomic spacing, so DM dominantly couples to collective modes rather than individual nuclei or electrons. These considerations have led theorists to consider a wide range of \textbf{new detection channels}, both in conventional and new materials.
There are already an impressive number of such proposed detection strategies in their early stages, many of which are funded at the prototype level, and some of which have already taken first data. Below we discuss a (non-exhaustive) list of examples; note that all of the following ideas are less than 6 years old. 

The first theorist-led collaboration which provided the motivation to push for thresholds well below 100 meV \cite{Hochberg:2015pha,Hochberg:2015fth} took advantage of an existing R\&D effort to push TESs to their fundamental sensitivity limit~\cite{Pyle:2015pya}. These papers proposed superconducting targets, with the gap as low as $\mathcal{O}$(meV) set by the Cooper pair breaking scale. The realization in~\cite{Hochberg:2015fth} (further studied in \cite{Gelmini:2020xir,Knapen:2021run,Hochberg:2021pkt}) that screening effects, as encapsulated by the dielectric function, could substantially limit the reach  led to consideration of narrow-gap insulators such as Dirac materials~\cite{Hochberg:2017wce}. In particular, condensed matter and materials science theorists were instrumental in suggesting the particular compounds considered in that paper, including ZrTe$_5$, showing the importance of collaboration between theorists in these different fields. The idea of DM-electron scattering with sub-eV thresholds was first implemented experimentally in  \cite{Hochberg:2021yud} using \textbf{superconducting nanowires} as both target and detector, based on a technique first proposed in~\cite{Hochberg:2019cyy}. In addition, \textbf{SPLENDOR}, an effort to use the narrow-gap semiconductors La$_3$Cd$_2$As$_6$ and Eu$_5$In$_2$Sb$_6$ read out via low-threshold cryogenic charge amplifiers~\cite{Phipps:2016gdx,Juillard:2019njs,LOI_CF1_CF2-IF1_IF2_Kurinsky-029}, has been funded for FY22-24 by the Los Alamos National Laboratory LDRD program. In this case, the collaboration was conceived after a particle theory seminar at LANL which was attended by members of the Materials Science Division, who suggested the novel materials that were first synthesized by that group \cite{rosa2020colossal}.

Along with with the developments in DM-electron scattering, polar materials such as GaAs and sapphire ($\text{Al}_2\text{O}_3$) were identified as excellent probes of DM-nuclear scattering via optical phonon excitation \cite{Knapen:2017ekk} because the dark photon mediator, which is the most plausible cosmologically \cite{Knapen:2017xzo} in the sub-MeV DM mass range, can easily excite optical phonon modes. This program was a theory/experiment joint venture from the very beginning and has led to the start of the \textbf{SPICE} collaboration. In parallel to the development of the solid state detectors, it has long been appreciated that superfluid helium has excellent potential as a detector for GeV-scale DM \cite{Guo:2013dt}. The \textbf{HeRALD} collaboration \cite{Hertel:2018aal,SPICEHeRALD:2021jba} was initiated to develop such a superfluid helium detector with meV energy thresholds, which suffices to detect DM as light as $\sim$1 MeV through nuclear recoils. Theory calculations of the relevant DM-phonon scattering processes~\cite{Schutz:2016tid,Knapen:2016cue,Acanfora:2019con,Caputo:2019cyg,Caputo:2020sys, Baym:2020uos} extended the reach of this experiment all the way down to DM as light as 10 keV, without the need to alter the design.
 HeRALD and SPICE will use much of the same sensor technology; their joint R\&D effort is carried out under the umbrella of the \textbf{TESSERACT} collaboration, which was awarded DMNI funding.


The concepts discussed so far all rely on calorimetry or charge collection. The idea of a \textbf{magnetic bubble chamber} \cite{Bunting:2017net} is orthogonal to these, and originated from a collaboration between theorists, experimentalists and chemists. This method envisions anti-aligning a large set of nano-scale magnets with an external magnetic field, such that a small energy deposit by the DM can trigger an avalanche reaction. The concept has been realized experimentally \cite{Chen:2020jia}, though so far still with a threshold well above what would be needed for a DM detector.


\section{Wave-like dark matter detection}
\label{sec:AxionDM}

As alluded to in Secs.~\ref{sec:Introduction}-\ref{sec:theory}, the connection between theory and experiment for axion DM searches is especially strong and goes back decades. Shortly after the proposal of the axion as a solution to the strong-CP problem via Peccei-Quinn (PQ) symmetry breaking \cite{Peccei:1977hh,Peccei:1977ur}, there were attempts to relate the axion decay constant $f_a$ to the electroweak scale \cite{Weinberg:1977ma,Wilczek:1977pj}; this, however, was quickly ruled out by beam dump searches, as the low decay constant yielded couplings that were too large.\footnote{Amusingly, the original E137 experiment \cite{Bjorken:1988as} at SLAC, which was designed to look for particles like the weak-scale axion, is continuing to pay dividends decades later by setting new limits on light dark-sector particles as described in Sec.~\ref{sec:Accelerator}; this illustrates the enormous return on investment provided by small-scale experiments!} Axions with higher $f_a$ were dubbed ``invisible'' as the extremely weak coupling was thought to be impossible to detect. Undeterred, a few years later Pierre Sikivie proposed two possible detection methods \cite{Sikivie:1983ip}: the ``helioscope'' for axions produced in the sun, and the ``haloscope'' for axions that could serve as a DM candidate and would thus be present in the laboratory in large occupation numbers. Both ideas led directly to a long-term program of funded experiments: the helioscope idea was taken up by \textbf{CAST} at CERN using a prototype LHC dipole magnet \cite{CAST:2020rlf}, and the haloscope idea was launched with the Rochester-Brookhaven-Fermilab collaboration in 1987, followed by an experiment at the University of Florida in 1990, which became the G2-funded flagship axion DM experiment \textbf{ADMX} \cite{ADMX:2021nhd}. Theoretical guidance was crucial for the success of ADMX, in particular the vision to see that ``invisible'' is only a relative term!

\subsection{Lumped-element searches for light axions}

The axion parameter space targeted by ADMX was in some sense pre-ordained by a numerical coincidence that related an experimental length scale to a natural theoretical target. In order to render the invisible axion visible, a large resonant enhancement of the electromagnetic signal from axion conversion is required, which is implemented through a high-$Q$ resonant cavity. The total signal power is proportional to the cavity volume, and the largest viable cavities have linear dimensions on the scale of meters, which corresponds to the Compton wavelength of an axion with mass $m_a \simeq 10^{-6} \ {\rm eV}$. This also happens to be within an order of magnitude or so of the axion mass expected if PQ symmetry is broken either after inflation, or before inflation with an order-1 misalignment angle in our Hubble patch. As we have emphasized in this whitepaper, over the decades since ADMX was first developed, the theoretical community has widened its scope beyond these naturalness arguments to ask what mass DM may possibly have, subject to as few theoretical prejudices as possible. In that context, the QCD axion may have a mass as low as $10^{-11} \ {\rm eV}$ for $f_a$ below the Planck scale, or a mass as large as $\sim 1 \ {\rm eV}$, which is still consistent with helioscope and other astrophysical constraints \cite{Irastorza:2018dyq}.

In the last decade, a renaissance of new theoretical ideas has given rise directly to a number of experiments probing the wider axion DM parameter space. One such idea is the use of ``lumped-element'' RLC circuits to provide the necessary resonance, thereby decoupling the physical size of the experiment from the resonance frequency and opening the possibility of a staged, scalable experimental program with successively larger $B$-field volumes. The first application of these ideas to axion detection was an unpublished conference talk suggesting the use of a tuneable LC circuit to enhance the axion-induced flux through a solenoidal magnetic field region \cite{ThomasTalk}, later fleshed out in \cite{Sikivie:2013laa}. A similar proposal for dark photon DM detection, \textbf{DM Radio}, also co-written by theorists and experimentalists, studied in detail many of the electrical engineering subtleties required to achieve thermal noise-limited sensitivity with a resonant readout circuit \cite{Chaudhuri:2014dla}. Inspired by these ideas, the \textbf{ABRACADABRA} proposal \citep{Kahn:2016aff} identified a toroidal geometry more robust to magnetic field noise and noted the advantages of broadband (as opposed to resonant) readout for a first pass through axion parameter space, which could still provide world-leading sensitivity. Due to a fortuitous synergy between the local expertise at MIT and the particular experimental needs for low-mass axion searches -- in particular, a neutrino physics group intimately familiar with cryogenics and a local small business that manufactured superconducting MRI magnets -- ABRACADABRA was able to jump off the journal page and into the real world, yielding the strongest constraints on neV-scale axions less than three years after the publication of the theory paper on which it was based \cite{Ouellet:2018beu,Ouellet:2019tlz}. 

Theorists continued as an integral part of the ABRACADABRA collaboration by proposing new statistical analysis techniques \cite{Foster:2017hbq} and implementing these techniques in the collection and analysis of the actual data; discussions between the DM Radio and ABRACADABRA collaborations also led to an ``optimality theorem'' quantifying the advantages of LC circuits for off-resonance broadband readout \cite{Chaudhuri:2018rqn,Chaudhuri:2019ntz}. New constraints in a similar parameter space were carved out shortly thereafter by the \textbf{ADMX-SLIC} collaboration \cite{Crisosto:2019fcj} (an implementation of Sikivie's solenoidal geometry) and the \textbf{SHAFT} collaboration \cite{Gramolin:2020ict} (implementing the ABRACADABRA geometry with a solid-state toroidal ferromagnet). As a result of all of this progress, the DM Radio collaboration (adapting their design for axion detection), joining forces with members of ABRACADABRA, was selected for funding under the DOE DMNI program in 2020. The eventual goal of the DM Radio program, \textbf{DM Radio-GUT}, is a 100 m$^3$ magnet that could probe the QCD axion with decay constants at the GUT scale~\cite{DMRadioGUT}. 
 As was the case with ADMX, theorists and theoretical guidance continue to be essential to ABRACADABRA/DM Radio as the experiments expand their reach in axion parameter space.

\subsection{Dish antennas and dielectric stacks for dark photons and axions\label{sec:antennas}}

While the lumped element RLC searches are probing wave-like DM in the mass range below $10^{-6}$ eV, the use of large spherical mirrors, so called ``dish antennas,"
may allow searches for axions above 50 $\mu$eV, beyond the reach of microcavity searches. The concept was first proposed by a collaborations of theorists and experimentalists about a decade ago~\cite{Horns:2012jf}. This method exploits the fact that reflective surfaces effectively convert oscillating axion-like or dark photon fields into electromagnetic radiation emitted perpendicular to the surface. Using a spherical surface, the emitted radiation is concentrated in the centre of the sphere where it can be detected. The technique is, in principle, broadband and allows exploration of a wide range of masses in a single measurement when appropriate detectors with a wide bandwidth are employed. Furthermore, the emitted power is proportional to the area of the surface and therefore easy to scale up. A dark photon search may be converted to an axion search with the addition of a magnetic field parallel to the surface of the converting mirror~\cite{Horns:2012jf}.

Three years after the proposal of the method, a group from Tokyo University published the first results from a dark photon DM search in the eV mass range with a dish antenna~\cite{Suzuki:2015sza}. Meanwhile, four further dish antenna experiments, exploiting various photon detectors sensitive in different frequency (mass) ranges, have published their results: \textbf{Tokyo-2}~\cite{Knirck:2018ojz} (with a theorist as lead author) around $10^{-3}$~eV, \textbf{Tokyo-3}~\cite{Tomita:2020usq} around $10^{-4}$~eV, \textbf{SHUKET}~\cite{Brun:2019kak} in Paris around $10^{-5}$~eV, 
and \textbf{FUNK}~\cite{FUNKExperiment:2020ofv} in Karlsruhe around 1~eV -- the last three experiments setting limits on the dark photon kinetic mixing parameter 
better or at least comparable to the strongest astrophysical constraint in the mass range accessible to their detectors. 

For the detection of axion-like DM, ideas for how to implement the required magnetic field have been put forward by the \textbf{BRASS}~\cite{BRASS:website} collaboration in Hamburg, which consists of theorists and experimentalists, and by the \textbf{BREAD}~\cite{BREAD:2021tpx}  collaboration led by Fermilab.  Theorists have also  developed a variant of the dish antenna haloscope that enhances the axion-photon conversion by exploiting many dielectric layers arranged to achieve constructive interference: the multilayer dielectric haloscope~\cite{Jaeckel:2013eha,Millar:2016cjp}. The \textbf{MADMAX} collaboration wants to exploit this method in order to 
search for DM axions in the mass range between 40 and 400 $\mu$eV~\cite{MADMAX:2019pub}, using a stack of 80 parallel dielectric disks with adjustable separations, which are placed in a strong magnetic field.  
A scaled-down prototype experiment is currently being set up in the Morpurgo magnet at CERN~\cite{MADMAX:2020ygz}. Note that the LAMPOST experiment discussed in Sec.~\ref{sec:eVDM} is essentially an implementation of this technique at the eV scale, a mass at which the distinction between particle-like and wave-like DM is blurred, illustrating the importance of cross-collaboration between these two direct detection communities.

\subsection{Axion-matter couplings}

While a large portion of the axion detection effort is focused on the coupling to photons, the only \emph{irreducible} coupling for the axion that solves the strong CP problem is to the QCD field strength, which becomes a coupling to nucleon electric dipole moments (EDMs) at low energies. This coupling is notoriously difficult to detect because the magnitude of the induced EDM is $4 \times 10^{-35} \ e \cdot {\rm cm}$, and thus some of the earliest detection proposals (initiated by theorists) relied on quantum sensing techniques \cite{Graham:2011qk}. A subsequent paper \cite{Graham:2013gfa} emphasized the importance of other axion-matter couplings, and a collaboration with experts in nuclear magnetic resonance (NMR) led to the \textbf{CASPEr} experiment \cite{Budker:2013hfa}. In particular, the effect of the axion ``wind'' coupling of the gradient of the axion field to spins is equivalent to that of an oscillating RF field with frequency $m_a$. CASPEr, which obtained initial funding from private foundations, has developed into two parallel experiments \cite{JacksonKimball:2017elr}, \textbf{CASPEr-Wind} (focusing on the axion gradient coupling to nuclear spins) and \textbf{CASPEr-Electric} (focusing on the EDM coupling). Preliminary data has been achieved by CASPEr-Wind at both ultra-low frequencies \cite{Garcon:2019inh} and by both CASPEr-Wind and CASPEr-Electric at frequencies in the range corresponding to an axion decay constant at the GUT scale \cite{Aybas:2021nvn}.

The idea behind the \textbf{QUAX} experiment is to exploit the effective magnetic field induced by the axion wind, but coupling to electron spins rather than nuclear spins \cite{Ruoso:2015ytk}. The ideas for this method to search for the axion date back to theory work from the 1980's \cite{Krauss:1985ph,Barbieri:1985cp,russians}. By tuning the Larmor frequency of the target to the axion mass, one can resonantly induce oscillations in the overall magnetization. QUAX is taking data and has already set constraints on the axion-electron coupling for a narrow mass range around $m_a\sim42\,\mu\text{eV}$ \cite{PhysRevLett.124.171801}.

The fact that the axion wind can be seen as an effective magnetic field allows the field of quantum magnetometry and quantum gyroscopes to be harnessed in the search for axion-like DM coupling to spins \cite{Safronova:2017xyt,Bloch:2019lcy}. These ultra-sensitive devices allow, among other things, extremely accurate measurements of anomalous magnetic fields, which affect either the nuclear spin or the electronic spin. Recently, theorists and experimentalists have joined forces to form the \textbf{NASDUCK} collaboration, which utilizes noble and alkali spin detectors in the search for DM. Since the fact that the DM mass is unknown requires high sensitivity at different frequencies, a wide array of experimental systems are being developed by the collaboration, with each system taking a distinct approach to maintain the required sensitivity in the relevant part of the spectrum.  Using a novel implementation of Floquet engineering, new world-leading constraints on ALP-neutron interactions have recently been obtained~\cite{Bloch:2021vnn}. In all of the experiments described in this subsection, theorists have been crucial in recognizing the analogies between DM detection and other quantum sensing modalities, and in identifying the ideal experimental collaborators from other fields of physics and chemistry.

\section{The Road Ahead}
\label{sec:RoadAhead}

To tell the stories of how theorists' work and new ideas have influenced the development of new  experiments, we have necessarily focused on several relatively mature ideas. Many of the experiments described above are at (or near) their final form, having taken preliminary data with full-scale experiments under construction, in advanced prototyping stages, or being designed.  Each experiment has reached this stage through a different path, including institutional and private-foundation support, small one-time support from national funding agencies, and in-kind support from experimentalists and theorists alike.  In many cases, the DOE's DMNI program has provided essential funds to bring these experiments to fruition or increase their scale; theorists have played an important dual role in both defining individual experiments and developing the overall case and priorities for a continuing small-experiment program in DM searches.  Though theorists remain passionately and essentially involved in these experiments, once the experimental concepts are clearly defined and motivated, theorists have typically transitioned back into more familiar roles: identifying new signals that can be probed, understanding backgrounds, and devising new high-level analysis paradigms. Meanwhile, the construction and operations of the experiments are led mainly by experimental collaborators. 

\subsection{New directions for the future}

We emphasize that \textbf{the development of new, innovative small-scale experiments by theorists is an ongoing story}.  Even as the experiments we describe above come of age, theorists continue to identify new directions for DM and weakly-coupled physics searches to branch out into.  These new proposals substantially extend and generalize their predecessors, or open up completely new directions. Many of these ideas have been submitted as Letters of Interest in the course of Snowmass 2021 \cite{LOI_CF1_CF2-IF1_IF2_Kurinsky-029,LOI_CF1_CF2-TF9_TF10-IF1_IF0_Asher_Berlin-049,LOI_CF1_CF0-IF1_IF2_Rouven_Essig-075,LOI_CF1_CF2-NF10_NF0-IF2_IF3_Kurinsky-101,LOI_CF1_CF2-IF1_IF2-102,LOI_CF1_CF0-TF9_TF0_Baxter-112,LOI_TF9_TF10-CF1_CF2_Zhengkang_Zhang-081,LOI_CF2-IF2-002,LOI_CF2_CF0-AF5_AF0_Kevin_Zhou-035,LOI_CF2_CF7-IF1_IF0_Oliver_Buchmueller-018,LOI_CF2_CF0-AF7_AF0-IF1_IF2-UF2_UF0_Jesse_Liu-179,LOI_CF2_CF0-RF3_RF0_Andrew_Geraci-130,LOI_CF2_CF0-RF3_RF0-IF1_IF0_Derek_F_Jackson_Kimball-103,LOI_CF2_CF0-TF9_TF0_Yonatan_Kahn-162,LOI_CF2_CF1-TF10_TF9-IF1_IF2_Lampost_Collaboration-131,LOI_RF6_RF0-EF10_EF0-CF1_CF0_Andrew_Whitbeck-111,LOI_RF6_RF0-NF2_NF3-AF2_AF5-099,LOI_RF6_RF0-NF3_NF0-AF5_AF0-084,LOI_RF6_RF0_Nhan_Tran-025,LOI_RF6_RF0_Torben_Ferber-020,LOI_RF6_RF2_Sean_Tulin-117,LOI_EF9_EF0-NF3_NF0-RF6_RF0_Matthew_Citron-072}. 





Recently proposed fixed-target experiments have had strong theory involvement, including a compact proton beam dump at Fermilab based on the SeaQuest experiment to search for long-lived particles \cite{LOI_RF6_RF0_Nhan_Tran-025}; proposals to search for dark matter scattering at $\sim 1-10$ GeV proton beam dumps at the FNAL complex \cite{LOI_RF6_RF0-NF2_NF3-AF2_AF5-099,LOI_RF6_RF0-NF3_NF0-AF5_AF0-084}; and using low-energy muon beams at Fermilab to search for muonic forces and dark matter \cite{Chen:2017awl,Kahn:2018cqs}. New ideas also exist to search for millicharged particles with the Fermilab and LHC accelerator facilities \cite{Foroughi-Abari:2020qar,Kelly:2018brz,LOI_EF9_EF0-NF3_NF0-RF6_RF0_Matthew_Citron-072}.  Theorists have also been closely involved in developing the science case for an $\eta$/$\eta^\prime$ factory \cite{LOI_RF6_RF2_Sean_Tulin-117}.
Specifically for long-lived particles, theorists have already studied the potential of dedicated detectors for long-lived particles at Belle II \cite{Dreyer:2021aqd}, as well as potential future colliders such as the FCC-ee \cite{Chrzaszcz:2020emg}, FCC-hh \cite{Bhattacherjee:2021rml}, the ILC \cite{Asai:2021ehn} and a future muon collider \cite{Cesarotti:2022ttv}. For the ILC, it was also pointed out that the intense gamma ray beam associated with undulator-based positron source can be used as a light-shining-through wall experiment for axion-like particles \cite{Fukuda:2022not}.

Examples of new directions led by theorists in just the past five years for particle-like direct detection are exploiting the anisotropies of condensed matter systems to allow directional detection of sub-GeV DM \cite{Essig:2011nj,Hochberg:2016ntt,Budnik:2017sbu,Cavoto:2017otc,Griffin:2018bjn,Coskuner:2019odd,Geilhufe:2019ndy,Blanco:2019lrf,Blanco:2021hlm,Coskuner:2021qxo} (leading to a daily modulation, which could greatly increase our ability to distinguish a DM signal from unknown backgrounds) and proposing new means to search for millicharge-like particle DM using macroscopic electromagnetic fields and high-$Q$ resonant detectors rather than scattering \cite{Berlin:2019uco,Berlin:2021kcm}. Proposed detectors inspired by advances in quantum sensing include the use of graphene Josephson junctions as detectors~\cite{Kim:2020bwm}. Theorists are also involved in the  DMSQUARE experiment~\cite{Avalos:2021fxm},  an effort to search for the diurnal modulation expected when sub-GeV DM interacts in the Earth, as well as motivating the design of a satellite-borne skipper-CCD detector to probe a subdominant DM component that has large interactions strengths~\cite{Emken:2019tni}. In addition, ion traps \cite{Budker:2021quh,Carney:2021irt} may have sensitivity to rather strongly interacting dark matter, while optical traps could be sensitive to very light dark matter \cite{Afek:2021vjy}. In both case, the dark matter particle in question ought to be subcomponent of the total dark matter density.

For wave-like DM, a number of new proposals have focused on the possibility of using interference effects to perform directional detection \cite{Knirck:2018knd,Foster:2020fln,Chen:2021bdr}. There have also been several new ideas to detect axions with novel condensed matter systems, including for axion-matter couplings \cite{Chigusa:2020gfs,Mitridate:2020kly,Chigusa:2021mci,Arvanitaki:2021wjk,Roising:2021lpv} (some of which revive ideas \cite{Barbieri:1985cp} first proposed by theorists around the same time as Sikivie's haloscope), and for the photon coupling at high frequencies by exploiting plasma resonances in metamaterials \cite{Lawson:2019brd} and axion quasiparticle modes in topological antiferromagnets \cite{Schutte-Engel:2021bqm}. Finally, there are still new ideas emerging from theorists for optimizing detection of the photon coupling in vacuum \cite{Berlin:2019ahk,Berlin:2020vrk,Lasenby:2019prg,Lasenby:2019hfz}, including heterodyne readout using superconducting radio-frequency (SRF) cavities which can achieve $Q$-factors of $10^{12}$ (see also Ref.~\cite{SnowmassTF10Cavities} for a detailed discussion of fundamental physics applications of SRF cavities).

Successfully and efficiently incubating these new ideas, as well as others yet to be devised, offers many benefits to the particle physics community.  Many draw on techniques and discoveries from other fields -- including, but not limited to, condensed matter physics, quantum sensing, quantum information science, chemistry, materials science, and accelerator physics -- enabling experimental particle physicists to capitalize on the successes of other disciplines, becoming experts in new technologies while interfacing with different research communities.  The most successful and multi-purpose techniques can be scaled up to larger-scale, pioneering experiments in different directions of exploration. 
Most importantly, such experiments open new search areas where a discovery --- though never guaranteed --- would truly transform the future of the field.  

At the same time, we wish to emphasize that \textbf{new theoretical ideas are still needed to delineate the space of allowed models for DM and dark sectors}. For example, the direct detection community has converged around a very small number of benchmark models -- including freeze-in through an ultralight kinetically-mixed dark photon, freeze-out through a heavy dark photon, and the QCD axion -- which satisfy the stringent constraints imposed by cosmology and astrophysics. However, these are almost certainly not the only models that exist in their respective mass ranges, and given how little is known about the dark sector from direct observation, the traditional role of phenomenologists as model-builders will be important in inspiring experiments to explore new parameter space.

\subsection{Challenges for the future}
To conclude, the experiments we highlighted here are only the first step towards a broader and more diversified experimental program in particle physics and we anticipate many more exciting innovations to come. However, the steady development of these ideas also faces significant challenges, including:  
\begin{itemize}
    \item \textbf{Network Inefficiency}: Coordination and knowledge-sharing between theorists and experimental experts is an absolute prerequisite to the development of new experiments, especially those based on a novel detection principle. It is often limited by the sizes of individual proponents' existing professional networks. ``Fortunate accidents'' are often needed for theorists to connect with experimentalists who not only have the requisite experimental expertise but also share an interest in the question at hand and are in a position to invest time in developing a new experiment; indeed, the stories we have told in the body of this whitepaper almost uniformly demonstrate the value of serendipity at key moments. \textbf{Recognizing the essential value of communication between theory and instrumentation  communities -- and between theorists of different disciplines -- and enabling this communication through both targeted workshops and opportunities for an unstructured exchange of ideas, is key to catalyzing the development of new experiments.} 
    \item \textbf{Funding}: Bottlenecks in funding and lack of clarity about how they should be funded can delay these small experiments by several years, roughly doubling their time to completion given their modest (by HEP standards) construction and data-taking requirements. This challenge has been especially acute when researchers not traditionally funded by HEP have key expertise, or when small ($<\$$2M) prototypes are insufficient to test a technique. \textbf{A stable, continuing source of funding for small, innovative experiments and associated R\&D would enable the realization of these ideas; the next generation of DM experiments may have its seeds in small-scale projects waiting for funding today.}
\end{itemize}
As it is less clear than ever where the next sign of new physics will appear first, it is critical for the future of the field that particle physics continue diversifying its portfolio of experiments. 
For this diversification to succeed, \textbf{a strong theory community is essential}, both to judiciously set priorities for parameter space targets, but also to spur new, potentially ground-breaking experimental concepts.

\clearpage

\bibliography{processed}

\end{document}